\documentclass[pdflatex,sn-mathphys-num]{sn-jnl}

\usepackage{comment}
\usepackage{graphicx}%
\usepackage{multirow}%
\usepackage{amsmath,amssymb,amsfonts}%
\usepackage{amsthm}%
\usepackage{mathrsfs}%
\usepackage[title]{appendix}%
\usepackage{xcolor}%
\usepackage{textcomp}%
\usepackage{manyfoot}%
\usepackage{booktabs}%
\usepackage{algorithm}%
\usepackage{algorithmicx}%
\usepackage{algpseudocode}%
\usepackage{listings}%
\usepackage{xcolor}
\usepackage{float}
\usepackage[most]{tcolorbox}
\usepackage{hyperref}
\usepackage{tabularx}
\usepackage{adjustbox}
\usepackage[table]{xcolor}

\makeatother
\usepackage{nicematrix}
\usetikzlibrary{fit}

\usepackage{makecell}   

\usepackage{booktabs,tabularx,array}

\usepackage[justification=justified,singlelinecheck=false]{caption}

\newcolumntype{Y}{>{\raggedleft\arraybackslash}X}

\newcommand{\reminder}[1]{\textcolor{red}{[REMINDER: #1]}}

\theoremstyle{thmstyleone}%
\newtheorem{theorem}{Theorem}
\newtheorem{proposition}[theorem]{Proposition}%

\theoremstyle{thmstyletwo}%
\newtheorem{example}{Example}%
\newtheorem{remark}{Remark}%

\theoremstyle{thmstylethree}%
\newtheorem{definition}{Definition}%

\begin{document}

\title[Article Title]{LEAD: LLM-enhanced Engine for Author Disambiguation}

\author[1,2]{\fnm{Giusy Giulia} \sur{Tuccari}}\email{giusy.tuccari@istc.cnr.it}
\equalcont{These are the leading authors of this work.}

\author[3]{\fnm{Lorenzo} \sur{Giammei}}\email{lorenzo.giammei@cnr.it}

\author[1]{\fnm{Andrea Giovanni} \sur{Nuzzolese}}\email{andreagiovanni.nuzzolese@cnr.it}

\author[1,2]{\fnm{Misael} \sur{Mongiovì}}\email{ misael.mongiovi@unict.it}

\author[3]{\fnm{Antonio} \sur{Zinilli}}\email{antonio.zinilli@cnr.it}

\author*[1]{\fnm{Francesco} \sur{Poggi}}\email{francesco.poggi@cnr.it}
\equalcont{These are the leading authors of this work.}

\affil*[1]{\orgdiv{Institute of Cognitive Sciences and Technologies (ISTC)}, \orgname{National Research Council of Italy (CNR)}, \orgaddress{
\country{Italy}}}

\affil[2]{\orgdiv{Department of Mathematics and Computer Science}, \orgname{University of
Catania}, \orgaddress{
\country{Italy}}}

\affil[3]{\orgdiv{Research Institute on Sustainable Economic Growth (IRCrES)}, \orgname{National
Research Council of Italy (CNR)}, \orgaddress{
\country{Italy}}}


\abstract{
Author Name Disambiguation (AND) is a long-standing challenge in bibliometrics and scientometrics, as name ambiguity undermines the accuracy of bibliographic databases and the reliability of research evaluation. 
This study addresses the problem of cross-source disambiguation by linking academic career records from CercaUniversità, the official registry of Italian academics, with author profiles in Scopus.
We introduce LEAD (LLM-enhanced Engine for Author Disambiguation), a novel hybrid framework that combines semantic features extracted through Large Language Models (LLMs) with structural evidence derived from co-authorship and citation networks. 
Using a gold standard of 606 ambiguous cases, 
we compare five methods: (i) Label Spreading on co-authorship networks; (ii) Bibliographic Coupling on citation networks; (iii) a standalone LLM-based approach; (iv) an LLM-enriched configuration; and (v) the proposed hybrid pipeline. 
LEAD achieves the best performance (F1 = 96.7\%, accuracy = 95.7\%) with lower computational cost than full LLM models.
Bibliographic Coupling emerges as the fastest and strongest single-source method. 
These findings demonstrate that integrating semantic and structural signals within a selective hybrid strategy offers a robust and scalable solution to cross-database author identification. 
Beyond the Italian case, this work highlights the potential of hybrid LLM-based methods to improve data quality and reliability in scientometric analyses.
}



\maketitle

\section{Introduction}\label{sec1}

Bibliographic databases play a central role in the research ecosystem. 
They are widely used to provide access to scientific publications, support the evaluation of academic performance, and guide the decision-making processes of funding agencies and academic institutions. 
The robustness of these databases, however, relies on the accurate identification of authors and their contributions.

Name ambiguity arises from several factors. In fact, different authors may share the same name (i.e., homonymy), while the same author may appear in multiple forms across publications due to spelling variations, transliterations, changes in affiliation, or incomplete metadata (i.e., synonymy). 
This issue affects not only the identification of the authors of published articles, but also of those cited within reference lists. 
As a result, it compromises the precision of bibliometric analyses, research evaluation exercises, and information retrieval tasks~\cite{smalheiser2009author,kim2014name,gomide2017name}.

The problem becomes even more pressing when multiple data sources are combined. 
Aligning bibliographic databases with other research information systems—such as those  tracking academic careers and institutional roles\cite{Martini20225633}, or databases covering patents \cite{Morrison2017}, research evaluation processes\cite{Bologna202319}, grant funding\cite{JACOB20111168} and doctoral training\cite{Zinilli2025}—requires reliable mechanisms for cross-database author identification. 
In the absence of robust disambiguation techniques, such integration risks producing fragmented or inaccurate representations of researchers’ outputs and trajectories, with potentially severe consequences for both scholarly and policy-oriented applications.
	
This task is known as {\em author name disambiguation (AND)}, which is a special case of {\em entity disambiguation} where entities are not labeled with unique identifiers\cite{dangelo2011}. 
The general problem has been extensively studied across several research communities, with methods designed for both structured and unstructured data. 
Within bibliometric applications, the problem arises in structured data environments, where articles are represented as records containing lists of authors together with attributes such as article titles, journal names, or publication years \cite{10.1162/qss_a_00081,Kim20212057}. 
Each mention of an author in these records constitutes an author instance that must be correctly assigned to the real-world individual it represents. 
Ensuring such mappings is essential for the integrity of bibliometric analyses and for the accuracy of the knowledge derived from them.

While bibliographic databases provide the most prominent context for AND, the challenge extends far beyond publications alone. 
Name ambiguity affects the integration of heterogeneous data sources that capture different facets of academic and professional activity. 
Examples include databases of individuals with biographical and career information, bibliometric databases of publications and citations, repositories of doctoral theses and academic habilitations, social networks, records of entrepreneurial activities and patents, and datasets documenting roles and appointments within institutions. 
In all these cases, entities of interest (i.e., people) are referenced primarily through names, often without persistent identifiers. 
As a result, multiple records referring to the same individual may appear fragmented across datasets, while homonyms may lead to false merges.
This issue is formally addressed within the research area of {\em entity resolution}~\cite{Christophides2021}, which is  the task of identifying and merging records that refer to the same real-world entity across multiple data sources in the absence of reliable unique identifiers to create a unified view. 
Author name disambiguation should be interpreted broadly, applying to all cases in which data sources concerning individuals need to be connected, reconciled, or integrated on the basis of personal names.
Whether the objective is to construct reliable publication profiles, to trace academic careers, to monitor the activities of academics within social networks, to link scientific output with patents and entrepreneurial ventures, or to support large-scale evaluations of research systems, the disambiguation of names is a prerequisite. 
Accordingly, bridging fragmented representations of people across diverse and partially overlapping information systems makes AND a cornerstone of contemporary bibliometrics, scientometrics, and research information management.

The importance of author name disambiguation has been recognized for decades. Early reflections on the problem date back to the 1960s and 1970s, when bibliometric studies began to expand, but systematic research efforts became more prominent in the 1990s and early 2000s with the large-scale digitization of publication records. Major commercial databases such as Web of Science and Scopus have long invested in proprietary author profiling systems, while more recent platforms such as Dimensions and OpenAlex continue to face challenges in providing comprehensive and accurate author-level data. In parallel, community-driven initiatives have sought to offer persistent digital identifiers, most notably ORCID, which provides a unique and portable identifier for researchers to mitigate ambiguity at its source.

Over the years, a wide spectrum of technical solutions has been proposed to address AND \cite{Hussain2017}. Initial approaches relied on heuristic and rule-based methods, using combinations of names, co-authors, and affiliations to distinguish between individuals \cite{caron2014,dangelo2011}. Later, supervised and unsupervised machine learning models were introduced, incorporating features such as citation patterns, venue information, and topical similarity \cite{Han2004296,Ferreira201215,10.1162/qss_a_00081}. More recently, graph-based algorithms and deep learning techniques have shown promise in scaling to massive datasets and capturing complex patterns of scholarly data\cite{mihaljevic2021,debonis2023,santini2022,Cappelli2025}. Large-scale projects such as ArnetMiner/AMiner~\cite{Tang2008990} and efforts within digital libraries like Microsoft Academic Graph~\cite{Wang2020396} have been instrumental in advancing the field and providing benchmarks for evaluation.

Despite these advances, author name disambiguation remains a complex and unresolved challenge. The continuous growth in scientific output, the variability of metadata quality across sources, and the need for interoperability among heterogeneous databases ensure that AND will remain a central concern for bibliometrics and research information management. Addressing this challenge is not merely a technical task, but a prerequisite for trustworthy bibliometric analyses and for ensuring the reliability of the data-driven processes that increasingly inform academic governance and science policy.

In this paper, we focus on a specific subproblem of author name disambiguation: establishing correspondences between records from CercaUniversità, a database containing career information of Italian academics from 2000 to the present, and the corresponding author profiles in Scopus. We introduce LEAD (LLM-enhanced Engine for Author Disambiguation), a novel hybrid approach that combines the strengths of large language models (LLMs) with network-based methods. LEAD is evaluated against three alternative approaches: (i) a standalone LLM-based model, (ii) the Citation-overlap method, which exploits citation networks, and (iii) the Label Spreading method, which leverages co-authorship networks. To assess the effectiveness of these methods, we constructed a hand-annotated gold standard consisting of 606 validated correspondences between CercaUniversità records and Scopus author profiles. Importantly, this dataset represents a “hard case” scenario, comprising highly ambiguous instances such as homonyms working at the same university, often in closely related disciplines. By tackling this challenging setting, our study provides insights into the strengths and limitations of different approaches to AND and highlights the potential of hybrid solutions in improving accuracy.

The paper is organized as follows. 
Section \ref{sec2} reviews related work, discussing existing approaches and highlighting current limitations. 
Section \ref{sec3} describes the materials and methods, including the construction of the dataset from Scopus and CercaUniversità and the disambiguation techniques under evaluation. 
Section \ref{sec4} presents the experimental results, comparing the performance of different methods through standard evaluation metrics. 
Finally, Section \ref{sec6} concludes the study by summarizing key contributions and suggesting directions for future research.


\section{Related work}\label{sec2}

Author name disambiguation (AND) is a widely investigated task that can be defined as a sub-task of entity resolution. 
More generally, AND aims to correctly assign publications to unique individuals, which involves handling name variations such as abbreviations, translation errors, pseudonyms, homonymy (different individuals sharing the same name), and synonymy (a single individual appearing under multiple name variants)\cite{binette}. Although numerous frameworks have been developed for the AND task, several limitations remain. As highlighted in a recent survey \cite{Cappelli2025}, a major challenge is the lack of high-quality annotated data, which is particularly problematic for deep learning approaches that require large and well-curated datasets. These models also face challenges related to heterogeneous metadata and class imbalance, which affect their generalizability.

Another key limitation is the lack of standardized benchmark datasets for consistent evaluation. Many existing datasets are manually curated, limited in scope or scale, and often affected by demographic biases, such as the underrepresentation of name variations across ethnic groups. These issues hinder fair comparisons among methods and underscore the need for more robust, diverse, and representative resources to support the development of scalable AND systems. The most common approach to solving the AND task consists of pre-processing author names into blocks \cite{on2005}, which group together all authors sharing the same last name and the first initial of their given name — a method known as the LN-FI (Last Name–First Initial) blocking. After blocking, a clustering algorithm is typically applied within each block to disambiguate authors based on similarity measures computed from metadata such as co-authors, affiliations, and titles.

In this work, we specifically address the problem of cross-source author name disambiguation, which goes beyond resolving homonymy and synonymy within a single dataset. Our goal is to disambiguate and align author identities across heterogeneous data sources by integrating them into a unified framework.

This task has been extensively studied, with proposed solutions generally falling into two categories: supervised and unsupervised \cite{Cappelli2025,debonis2023}.  
While the former require substantial labeled data, often obtained through labor-intensive manual annotation, the latter are designed to operate without any reliance on human-labeled input. 

Supervised approaches typically rely on machine learning algorithms, including classical models such as Random Forest, Logistic Regression, Gradient Boosting, and Naive Bayes \cite{rehs2021, kim2021, mihaljevic2021, Zinilli2025}, as well as more recent deep learning frameworks \cite{chen2022}.
Rehs \cite{rehs2021} proposed a supervised machine learning approach for author name disambiguation in the Web of Science (WoS) dataset, comparing the performance of Random Forest and Logistic Regression classifiers. The resulting pairwise predictions are then aggregated using a graph-based post-estimation strategy based on the Infomap community detection algorithm \cite{rosvall2008}. The approach leverages a diverse set of features derived from metadata, including country and institution, topic modeling, subject classifications, keywords, and name-based statistics.
Mihaljević et al.\cite{mihaljevic2021} addressed the author name disambiguation problem in the SAO/NASA Astrophysics Data System (ADS) dataset using a supervised learning approach combined with graph-based techniques. They compare three different classifiers (Decision Tree, Random Forest, and Histogram-based Gradient Boosting Decision Tree) to estimate the likelihood that pairs of author records refer to the same individual, based on features extracted from publication metadata. These pairwise predictions are then used to construct a weighted graph, where nodes represent author entities and edges encode the computed similarity scores. A label propagation algorithm is applied to this graph to identify clusters corresponding to distinct authors.
Chen et al.\cite{chen2022} proposed an end-to-end framework for author name disambiguation in large-scale academic systems, addressing the on-the-fly disambiguation problem, in which newly arriving publications must be continuously assigned to existing or new author profiles. The proposed method, CONNA, was evaluated on the OAG-WhoIsWho and KDD Cup 2013 datasets. It integrates a matching component and a decision component, which determines whether to assign the target paper to the top-ranked candidate or to create a new author identity (NIL). Both components are jointly trained via reinforcement learning, allowing for iterative correction of their respective errors. However, this technique substantially differs from our objective, as it targets dynamic, real-time disambiguation and relies heavily on labeled data, whereas our approach is unsupervised and operates in a static evaluation setting.

While most supervised methods rely on structured features or deep architectures trained on labeled data, recent work has begun to explore the use of Large Language Models (LLMs) to incorporate semantic information from textual metadata. Zhao et al.\cite{zhao2024} introduced a search-enhanced LLM framework that retrieves web content—including non-English sources—to build more comprehensive author profiles, improving disambiguation in multilingual settings. Their method combines name translation, web search, and profile comparison modules orchestrated by LLM-based agents. However, such solutions often operate in real-time settings, relying on external resources and significant computational power, and may struggle to scale to large static corpora due to LLM context length limitations.
In contrast, our approach integrates LLM-based embeddings into a fully unsupervised, graph-based framework, enabling scalable and language-agnostic disambiguation without supervision or external search tools.

 LAND, proposed by Santini et al.\cite{santini2022}, is an unsupervised framework for author name disambiguation in scholarly knowledge graphs. It learns entity representations by combining structural information from graph triples—using the DistMult model—with literal attributes such as publication titles and years. Literal information is integrated through LiteralE, leveraging SPECTER embeddings for textual data. The framework applies a blocking strategy based on last name and first initial, followed by hierarchical agglomerative clustering.
 Cheng et al.\cite{cheng2024} introduced BOND, a framework that jointly models pairwise similarity and clustering for author name disambiguation. It adopts a self-bootstrapping strategy, where a multi-task neural network estimates local similarities, which are then clustered to produce pseudo-labels. These labels are iteratively fed back to refine the similarity model. BOND requires no manually labeled data and is extended in BOND+ with ensemble clustering and post-processing modules.
Luo et al. \cite{luo2022} introduced a novel approach, namely NDCC, that model bibliographical data as a heterogeneous multipartite network and an iterative method via collective clustering.
GRAND\cite{huang2025} modeled author mentions within a heterogeneous network by leveraging both structural and semantic information. It adopts a meta-path-based random walk strategy to explore meaningful relational patterns and employs a heterogeneous skip-gram model to learn node embeddings from the resulting vertex sequences. By aggregating contextual signals from both the structural (e.g., co-authorship, venue) and semantic (e.g., titles, abstracts, keywords) neighborhoods, GRAND generates author representations that integrate multiple information sources without relying on manually crafted features.
In \cite{dangelo2020}, the authors proposed a cluster-based approach that combines two existing techniques, namely the CvE method \cite{caron2014} and the DGA method \cite{dangelo2011}. This approach follows the three-step procedure of \cite{caron2014} and includes a validation step inspired by \cite{dangelo2011} to perform author name disambiguation.
Firstly, a pre-processing phase is carried out by creating author name blocks \cite{on2005}. This step identifies potential candidate records and significantly reduces the computational cost of the subsequent phases.
Secondly, publication–author combinations (PACs) are generated, and a rule-based scoring system is applied. The scoring is based on various types of rules, including author metadata, source information, and citation relations.
Subsequently, in the post-processing phase, candidate author oeuvres are merged if they share the same email address. Finally, the resulting clusters are validated and refined through the use of an external institutional database and additional filtering rules.
This study is the closest to our work, as it also makes use of institutional information. However, in their approach institutional data are employed only in a post-processing validation step to refine clusters derived from bibliometric records, rather than being directly incorporated into the disambiguation process. In contrast, our method integrates heterogeneous data sources from the outset within a unified framework for cross-source author name disambiguation.


\section{Materials and methods}\label{sec3}
Our goal is to identify matching authors from two different sources in the absence of a shared author identifier. Unlike other approaches, we do not assume that authors are unique within each source -- i.e., an author in one source may correspond to more than one entry in the other source --, nor that all authors from one source are present in the other one -- i.e., given an author in one source, it is not sufficient to simply rank the corresponding entries in the other source; we must also decide whether the best-ranked author is a suitable match or not.

Next, we present the use-case benchmark dataset and the methodologies evaluated to address the disambiguation task. These include: (i) a large language model (LLM)-based approach; (ii) two graph-based methods, namely the Citation-overlap method, which leverages a citation network, and the Label Spreading method, which relies on a co-authorship network; and (iii) a hybrid approach that combines these techniques.
The use-case benchmark dataset can be found in~\cite{goldstandard}, and the software developed for our work is provided in a dedicated GitHub repository.\footnote{\url{https://github.com/fossr-project/LEAD}}.

\subsection{Dataset description}\label{subsec2}

The dataset used in our study was specifically designed as a list of hard pairs for the author name disambiguation task. It was built by combining and processing data from two sources: a
bibliographic source (i.e., Scopus\footnote{\url{https://www.scopus.com}}) and institutional source (i.e., CercaUniversità\footnote{\url{https://cercauniversita.mur.gov.it/} is a service provided by the Italian Ministry of University and Research (MUR) that provides information and statics about Italian professors, universities, degree programs, students, fundings, etc.}).
Scopus provides bibliographic metadata and citation information, while CercaUniversità offers official records on the academic careers of Italian researchers and professors.


Before delving into the details of how the dataset was created, it is important to provide some background on the organization of the Italian academic system.  In Italy each academic is bound to a specific Recruitment Field (RF), which corresponds to a scientific field of study. 
RFs are organized in groups called Recruitment Field Groups (RFGs), 
which are in turn sorted into 14 Scientific Areas (SAs) organized in a taxonomy\footnote{Please refer to the Italian Ministerial Decree 855/2015 ``Redefinition of Academic Recruitment Fields''~\cite{DM855}, which provides the legal framework of the Italian academic system.}. 
In this taxonomy each of the 190 RFs is identified by an alphanumeric code in the form AA/GF, where AA is the ID of the SA (in the range 01-14), G is a single letter identifying the RFG, and F is a digit denoting the RF. 
For example, the code of the RF ``Financial Markets and Institutions'' is 13/B4, which belongs to the RFG ``Business administration and Management'' (13/B), which is part of the SA ``Economics and Statistics'' (13). 
The 190 RFs are further specified in 370 Academic Disciplines (ADs), which aim to precisely classify the teaching and scientific activities of Italian professors and researchers.
The 14 SAs are listed in Table~\ref{scientificareas}, together with the division into RFGs, RFs and ADs\footnote{The complete list of the 190 RFs is available at \url{https://www.cun.it/uploads/storico/macrosettori_concorsuali_english.pdf}; and the list of the 370 ADs is available at \url{https://www.cun.it/uploads/storico/settori_scientifico_disciplinari_english.pdf}}. 


\begin{table}[]
\centering
\caption{The 14 Italian SAs. For each we report the numeric ID, a three-letter code, the name of the area and the number of RFGs, RFs and ADs it contains.}
\label{scientificareas}
\renewcommand{\arraystretch}{1.3}

\begin{tabularx}{\textwidth}{cllccc}
\toprule
\textbf{\begin{tabular}[c]{@{}c@{}}Area\\ID\end{tabular}} & \textbf{\begin{tabular}[c]{@{}c@{}}Area\\Code\end{tabular}} & \textbf{\begin{tabular}[c]{@{}c@{}}Area\\Name\end{tabular}} & \textbf{\begin{tabular}[c]{@{}c@{}}N. of\\RFGs\end{tabular}} & \textbf{\begin{tabular}[c]{@{}c@{}}N. of\\RFs\end{tabular}} & \textbf{\begin{tabular}[c]{@{}c@{}}N. of\\ADs\end{tabular}} \\
\midrule
01 & MCS & Mathematics and Informatics & \phantom{00}2 & \phantom{00}7 & \phantom{0}10  \\ 
02 & PHY & Physics & \phantom{00}3 & \phantom{00}6 & \phantom{00}8  \\
03 & CHE & Chemistry & \phantom{00}4 & \phantom{00}8 & \phantom{0}12 \\ 
04 & EAS & Earth Sciences & \phantom{00}1 & \phantom{00}4 & \phantom{0}12 \\ 
05 & BIO & Biology & \phantom{0}10 & \phantom{0}14 & \phantom{0}19 \\ 
06 & MED & Medicine & \phantom{0}10 & \phantom{0}27 & \phantom{0}50 \\ 
07 & AVM & Agricultural and veterinary sciences & \phantom{00}7 & \phantom{0}14 & \phantom{0}30 \\ 
08 & CEA & Civil Engineering and Architecture & \phantom{00}5 & \phantom{0}12 & \phantom{0}22 \\ 
09 & IIE & Industrial and Information Engineering & \phantom{00}7 & \phantom{0}21 & \phantom{0}42 \\ 
10 & APL & Antiquities,Philology,Literary Studies, Art History & \phantom{0}11 & \phantom{0}21 & \phantom{0}77 \\
11 & HPP & History, Philosophy, Pedagogy and Psychology & \phantom{00}4 & \phantom{0}18 & \phantom{0}34 \\ 
12 & LAW & Law Studies & \phantom{00}7 & \phantom{0}16 & \phantom{0}21 \\ 
13 & ECS & Economics and Statistics & \phantom{00}3 & \phantom{0}15 & \phantom{0}19 \\ 
14 & PSS & Political and Social Sciences & \phantom{00}3 & \phantom{00}7 & \phantom{0}14 \\
\midrule
 & \textbf{Total} & & \textbf{\phantom{0}77} & \textbf{\phantom{}190} & \textbf{\phantom{}370} \\
\bottomrule
\end{tabularx}
\end{table}


As mentioned previously, CercaUniversità provides information on Italian professors and researchers from 2000 to the present. 
For example, Table~\ref{recordCercauniversita} shows the records of three namesakes serving in the Italian Academy as of 31 December 2024. 
We observe that two of them are employed at the University of Bologna and work in rather similar fields (namely computer science and electronic engineering), while the third is a researcher in the field of legal history at the University of Trieste.

\begin{table}[]
\caption{Three example records of namesakes from CercaUniversità for the year 2024.}
\label{recordCercauniversita}
\small
\setlength{\tabcolsep}{1.2pt}
\footnotesize
\begin{tabular}{|l|l|c|c|l|l|l|l|}
\hline
\multicolumn{1}{|c|}{\textbf{Year}} & \multicolumn{1}{c|}{\textbf{Role}} & \textbf{Name} & \textbf{Gender} & \multicolumn{1}{c|}{\textbf{\begin{tabular}[c]{@{}c@{}}Recruitment\\Field (RF)\end{tabular}}} & \multicolumn{1}{c|}{\textbf{\begin{tabular}[c]{@{}c@{}}Academic\\Discipline (AD)\end{tabular}}} & \multicolumn{1}{c|}{\textbf{University}} & \multicolumn{1}{c|}{\textbf{Department}} \\ \hline
2024 & \begin{tabular}[c]{@{}l@{}}Associate\\ Professor\end{tabular} & \begin{tabular}[c]{@{}l@{}}Rossi \\Davide\end{tabular} & Male   & \begin{tabular}[c]{@{}l@{}}INF/01\\ (Informatics)\end{tabular}                          & \begin{tabular}[c]{@{}l@{}}01/B1\\ (Informatics)\end{tabular}                          & Bologna    & \begin{tabular}[c]{@{}l@{}}Computer Science \\and Engineering\end{tabular} \\ \hline
2024 & \begin{tabular}[c]{@{}l@{}}Associate\\ Professor\end{tabular} & \begin{tabular}[c]{@{}l@{}}Rossi \\Davide\end{tabular} & Male   & \begin{tabular}[c]{@{}l@{}}ING-INF/01\\ (Electronic \\Engineering)\end{tabular}           & \begin{tabular}[c]{@{}l@{}}09/E3\\ (Electronics)\end{tabular}                          & Bologna    & \begin{tabular}[c]{@{}l@{}}Electrical, Electronic and \\Information Engineering \\"Guglielmo Marconi"\end{tabular} \\ \hline
2024 & Researcher                                                    & \begin{tabular}[c]{@{}l@{}}Rossi \\Davide\end{tabular} & Male   & \begin{tabular}[c]{@{}l@{}}IUS/19\\ (History of \\Medieval and \\Modern Law)\end{tabular} & \begin{tabular}[c]{@{}l@{}}12/H2\\ (History of \\Medieval and \\Modern Law)\end{tabular} & Trieste    & \begin{tabular}[c]{@{}l@{}}Legal, Language,\\ Interpreting\\ and Translation Studies\end{tabular}                  \\ \hline
\end{tabular}
\end{table}

The aim of this work is to link each record from CercaUniversità to the correct profile(s) in Scopus, so as to have for each academic both career information within the Italian university system and information on scientific output.
Each academic may have no match in Scopus (in cases where they have no publications indexed by Scopus), a single match, or more than one. 
The 
case of an author with multiple profiles in Scopus is common, and typically occurs when an author has changed research field, or when he/she has had a very long career and some minor publications—e.g., book prefaces, editorials, etc.—have not been correctly associated with the author’s main profile.

CercaUniversità provides a partial solution to this problem. Since 2022, in fact, CercaUniversità has been providing the Scopus Author IDs (AUIDs, i.e., the unique identifiers assigned by Scopus to each author in its database) of Italian academics who are members of a doctoral board at an Italian university.
For the year 2022, for instance, CercaUniversità provides the Scopus AUIDs of 21,476 Italian academics out of a total of 61,288 (35.04\%).

To create a gold standard 
for evaluating the effectiveness of the disambiguation methods presented in this work (see Section \ref{subsec3}), we proceeded as follows. 
For each CercaUniversità profile for which we do not have the corresponding Scopus AUID, an initial string-matching phase was performed using the Scopus Search API\footnote{Scopus Search API documented at \url{https://dev.elsevier.com/documentation/ScopusSearchAPI.wadl}} to search for candidate Scopus matches based on four fields: first name, last name, affiliation city and the year of interest.
Another field provided by CercaUniversità that we tested for this purpose was the affiliation. However, after running several tests, we decided not to use it because the Scopus API is not reliable in handling the numerous variations in department names (e.g., Italian vs. English, full vs. abbreviated vs. acronym, etc.).
The results of this step for each of the 39,812 records relating to the year 2022 for which CercaUniversità does not provide the corresponding Scopus AUID are as follows: for 16,289 (40.91\%) records the Scopus Search API returned zero AUIDs, for 23,299 (58.53\%) records a single AUID, and for the remaining 224 (0.56\%) records two or more AUIDs.



\begin{figure}
    \centering
    \includegraphics[width=1\linewidth]{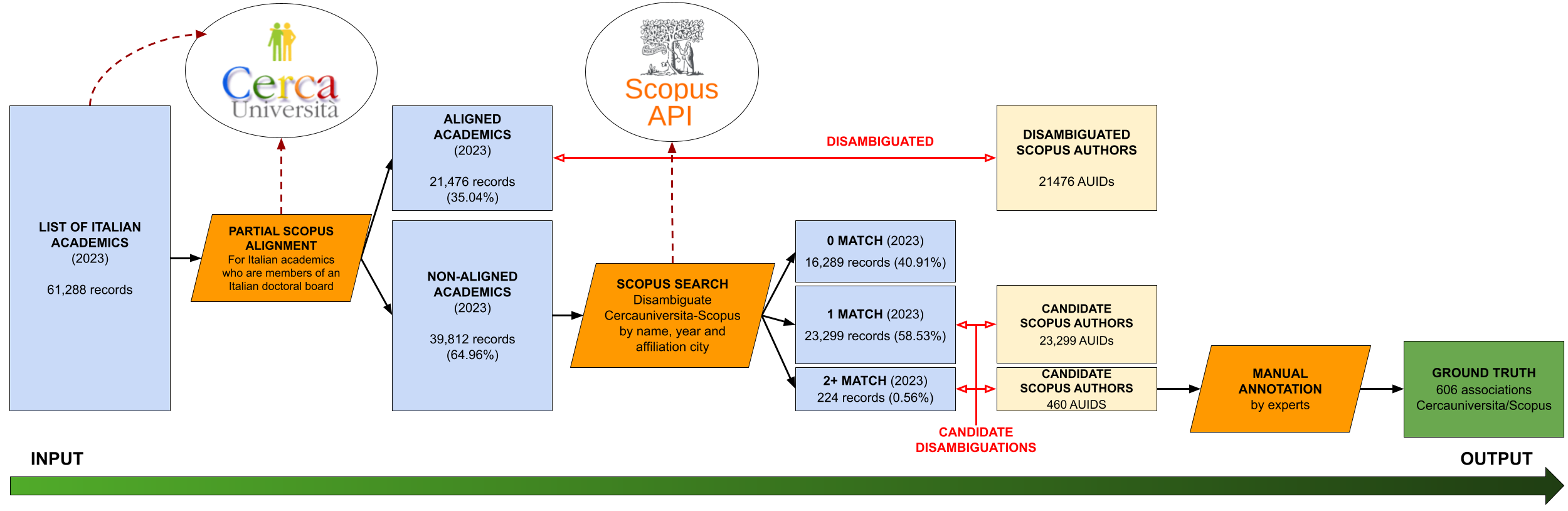}
    \caption{Processing workflow to generate the ground truth.}
    \label{fig:workflow1}
\end{figure}

To thoroughly stress the disambiguation systems and better assess their effectiveness, we created an {\em ``hard dataset''} based on highly complex cases involving multiple candidates. For example, we selected cases of homonyms working in similar disciplines, authors with multiple Scopus profiles, and so on. 
The resulting dataset was manually annotated and verified to serve as ground truth for evaluating the disambiguation methods presented in this study.
It is important to note that a single academic from CercaUniversità may be associated with multiple candidate  Scopus Author IDs (AUIDs), each evaluated independently.
This depends on the initial string-matching phase, which often returns more than one potential author profile due to name ambiguities. Additional factors may include human error during profile creation or inconsistent data entry by the authors themselves. The main steps of the process that led to the creation of the ground truth are shown in Figure~\ref{fig:workflow1}. 

The final gold standard 
is available on Zenodo\footnote{\url{https://doi.org/10.5281/zenodo.17244811}}~\cite{goldstandard}
and consists of 606 entries, each corresponding to a candidate association between a CercaUniversità record and a Scopus author AUID to be evaluated. 
The dataset is fairly balanced and contains 394 correct associations (65\%) and 212 incorrect ones (35\%).
As shown in Table~\ref{tab:groundtruth}, each entry includes personal and institutional information (name, surname, RF, AD, and university)
, the candidate AUID, and a binary label (correct) indicating whether the candidate was manually validated as the true author profile. 
Table~\ref{tab:area-distribution} reports the number of unique researchers in the gold standard grouped by Italian SAs. 
The distribution is highly unbalanced, with SA 06 (Medicine) and SA 09 (Industrial and Information Engineering) being the most represented. Some areas, such as SA 10 (Humanities) and SA 14 (Social Sciences), include very few entries. This disparity may reflect both the underlying composition of the ground truth and structural differences in publication patterns across disciplines, particularly regarding coverage in Scopus-indexed venues.

\begin{table}[htbp]
\caption{Structure of the manually annotated gold standard dataset used for evaluation.}
\label{tab:groundtruth}
\centering
\small
\setlength{\tabcolsep}{1pt} 
\begin{tabular}{|l|p{10cm}|}
\hline
\textbf{Field} & \textbf{Description} \\
\hline
\texttt{first\_name}, \texttt{last\_name} & First and last name of the academic \\
\texttt{RF} & Recruitment Field code \\
\texttt{AD} & Academic Disciplines code \\
\texttt{university} & University of affiliation \\
\texttt{auid} & Candidate Scopus Author ID \\
\texttt{correct} & Equal to 1 if the AUID was manually validated as correct, 0 otherwise \\
\hline
\end{tabular}
\end{table}




In addition to the gold standard, in this work we used publication data from (i) academics for whom CercaUniversità provides a Scopus AUID (e.g., 21,476 Italian academics who in 2023 were members of a doctoral board), and (ii) the 606 Scopus profiles included in the ground truth, whose correspondence with CercaUniversità was evaluated by the disambiguation methods.

In particular, for each Scopus author profile, we used the Scopus Author Retrieval API\footnote{Scopus Author Retrieval documented at \url{https://dev.elsevier.com/documentation/AuthorRetrievalAPI.wadl}} to download the metadata associated with the author, the aforementioned Scopus Search API to retrieve publication metadata since 2000, and the Scopus Abstract Retrieval API\footnote{Scopus Abstract API documented at \url{https://dev.elsevier.com/sc_abstract_retrieval_views.html}} to obtain the reference lists of these publications.

This dataset enabled the evaluation of multiple disambiguation techniques, that is: (i) an LLM-based method using titles, keywords, and abstracts (Section \ref{subsubsec1}); (ii) a Label Spreading approach exploiting co-author networks (Section \ref{subsec4}); (iii) a citation-network-based method (Section \ref{subsec5}); and (iv) a hybrid 
approach combining LLM and citation network analysis (Section \ref{subsec6}).

\begin{table}[h]
\centering
\caption{Distribution by Italian SAs of unique researchers in the gold standard.}
\label{tab:area-distribution}
\begin{tabular}{|c|l|r|}
\hline
\textbf{Area} & \textbf{Disciplinary Area} & \textbf{Entries} \\
\hline
01 & Mathematics and Computer Science      & 25 \\
02 & Physics                               & 20 \\
03 & Chemistry                             & 17 \\
04 & Earth Sciences                        & 10 \\
05 & Biology                               & 29 \\
06 & Medicine                              & 91 \\
07 & Agricultural and Veterinary Sciences  & 16 \\
08 & Civil Engineering and Architecture    & 16 \\
09 & Industrial and Information Engineering& 61 \\
10 & Antiquities, Philology, Literary Studies, Art History & 0 \\
11 & History, Philosophy, Pedagogy and Psychology & 7 \\
12 & Law Studies                           & 1 \\
13 & Economics and Statistics              & 9 \\
14 & Political and Social Sciences         & 1 \\
\hline
\end{tabular}
\end{table}

\subsection{Disambiguation Approaches}\label{subsec3}

In this section, we present the disambiguation approaches evaluated in this study.
Section~\ref{subsec4} introduces the Bibliographic Coupling approach, based on shared references in the citation network, while Section~\ref{subsec5} describes the Label Spreading method, which leverages information from the co-authorship network.
Section~\ref{subsubsec1} details the LLM-based module, which exploits textual metadata to distinguish between candidate authors.
Building on these individual components, Section~\ref{subsec6} presents the LLM-enriched variants, which integrate information from Label Spreading and Bibliographic Coupling into the LLM-based module.
Finally, Section~\ref{subsec7} outlines the proposed hybrid method, LEAD, which selectively combines the strengths of the previous approaches. 
The approaches presented were implemented in Python, and the corresponding code is publicly available in a dedicated GitHub repository\footnote{\url{https://github.com/fossr-project/LEAD}}.

\subsubsection{Bibliographic Coupling disambiguation}\label{subsec4}

A wide range of approaches to author name disambiguation leverage the structural properties of citation networks \cite{Ferreira201215,Hussain2017,Rodrigues2024765}. 
These methods exploit the intuition that scholarly identity can be inferred from patterns of citation and co-citation. 
Among them, \textit{bibliographic coupling} (also known as citation overlap) has emerged as particularly relevant for our task \cite{Maltseva2024110}. 
Bibliographic coupling is a method for assessing the similarity between two source documents by identifying the number of shared bibliographic references \cite{kessler1963bibliographic}.
When documents share one or more previously published bibliographic references, they are considered bibliographically coupled. 
The strength of this coupling is determined by the number of references they have in common, the more shared references, the stronger the connection. 
By measuring the degree of similarity between authors based on the overlap of their cited references, bibliographic coupling provides a robust indicator of intellectual proximity, indicating a probability of them addressing similar subject matters.
This property makes it a valuable tool for distinguishing between homonymous researchers, especially in domains where textual features or affiliation data alone may be insufficient. 
In contrast, co-citation analysis \cite{willems1969sociological,Bayer1990444} measures how often two documents are cited together in subsequent publications. 
As Garfield (2001) notes, ``Bibliographic coupling is retrospective whereas co-citation is essentially a forward-looking perspective''~\cite{garfield2001from}.

Bibliographic coupling presents distinctive characteristics that make it particularly well-suited for the objectives of this study. 
Specifically, when considering an author profile from the Italian ``CercaUniversità'' database it becomes necessary to establish a reliable correspondence with the author’s Scopus profile. 
The latter includes comprehensive bibliometric metadata and citation-based information. 
As previously outlined, the candidate correspondence was derived through a textual matching procedure. 
This approach relies on the alignment of key attributes, namely the author's first name, surname, city of affiliation, and the relevant year under consideration. 
Such criteria provide a preliminary, yet essential basis for establishing potential matches across datasets, but a subsequent verification and refinement of author identification is a crucial step for the validation and enrichment of the integrated dataset, thereby enabling more robust analyses of scholarly activity and academic trajectories.

The procedure adopted in this study can be summarized as follows. 
For each of the 190 RFs, we retrieved the Scopus Author IDs of those Italian scholars for whom the CercaUniversità platform provides an alignment with Scopus (see Section \ref{subsec2}). 
For these authors, we collected their publications and extracted the set of cited works from the reference sections. 
In this way, we constructed, for each discipline, a corpus of articles cited by Italian academics. 
Subsequently, for each candidate profile to be verified (identified by a Scopus Author ID), we applied the same process to generate the author's individual set of cited references. 
Since the match to be validated links a Scopus author with an Italian academic associated with a specific RF, 
we then assessed the overlap between the candidate's references and those cited by Italian scholars in the corresponding RF. 
To assess the robustness of this strategy across different time spans, we further compared the results obtained when considering common references derived from publications within a shorter 4-year interval (2020–2023) and a longer 8-year interval (2016–2023).
This approach provides a systematic and citation-based framework for validating candidate correspondences across datasets.

To illustrate the methodology employed, we provide an example drawn from the gold standard. Specifically, we consider the case of verifying two potential Scopus matches for Associate Professor Davide Rossi, who in 2022 was affiliated with the University of Bologna in the RF 09/E3 (i.e., Electronic Engineering). 
Table \ref{tab:bibliocoupling01} presents the details of the two candidate records, showing information about the academics in CercaUniversit\`{a} and the corresponding candidate profiles. 
Table~\ref{tab:bibliocoupling02} reports the details of the results obtained using the bibliographic coupling–based method for the two records under examination. The analysis reveals a substantial difference in the degree of overlap: for the first record, there is a 92.2\% overlap (2,206 shared references out of 2,393 between the candidate author and the 271 scholars in the 09/E3 sector), while for the second record the overlap amounts to only 5.7\% (27 shared references out of 470 in total). A subsequent manual verification confirmed that the first correspondence is correct, while the second is not. In fact, the latter Scopus profile belongs to a homonymous associate professor at the same university but in the RF 01/B1 (i.e., Informatics). 
This example highlights the discriminative power of bibliographic coupling in author disambiguation within the Italian academic system.


\begin{table}[h!]
\centering
\caption{An example of two records in the gold standard, containing the data of a university professor from Bologna, and two possible corresponding Scopus profiles (based on the fields first name, last name, and affiliation city) to be validated.}
\label{tab:bibliocoupling01}
\renewcommand{\arraystretch}{2}
\setlength{\tabcolsep}{2.5pt}
\begin{tabular}{@{}ccccccccc@{}}
\toprule
& \multicolumn{4}{c}{\textbf{CERCAUNIVERSIT\`{A}}} & \multicolumn{4}{c}{\textbf{CANDIDATE AUTHOR IN SCOPUS}} \\ 
\cmidrule(lr){2-5} \cmidrule(lr){6-9}
\makecell{\textbf{Record}\\\textbf{ID}} & \makecell{\textbf{Given}\\\textbf{Name}} & \textbf{Surname} & \textbf{RF} & \textbf{University} 
 & \makecell{\textbf{Scopus}\\\textbf{AUID}} & \makecell{\textbf{Given}\\\textbf{Name}} & \textbf{Surname} & \makecell{\textbf{Affiliation}\\\textbf{City}} \\ 
\midrule
14 & DAVIDE & ROSSI & 09/E3 & Bologna & 7103169675 & DAVIDE & ROSSI & Bologna \\
15 & DAVIDE & ROSSI & 09/E3 & Bologna & 57194011914 & DAVIDE & ROSSI & Bologna \\
\bottomrule
\end{tabular}
\end{table}


\begin{table}[h!]
\centering
\renewcommand{\arraystretch}{2}
\setlength{\tabcolsep}{4pt}
\caption{Publication and reference data for (1) the 271 academics in sector 09/E3 for whom CercaUniversit\`{a} provides an alignment with Scopus, and (2) the two candidate Scopus profiles to be disambiguated. In addition, information is reported on the overlapping references between these two entities. The data were obtained considering an eight-year period from 2016 to 2023.}
\label{tab:bibliocoupling02}
\begin{tabular}{cccccccc}
\toprule
 & 
\multicolumn{3}{c}{\makecell{\textbf{AUTHORS IN}\\\textbf{RF 09/E3}}} & 
\multicolumn{2}{c}{\makecell{\textbf{CANDIDATE}\\\textbf{AUTHOR}}} & 
\multicolumn{2}{c}{\textbf{RESULTS}} \\
\cmidrule(lr){2-4} \cmidrule(lr){5-6} \cmidrule(lr){7-8}
\makecell{\textbf{Record}\\\textbf{ID}} & \makecell{\textbf{\# of}\\\textbf{authors}} & \makecell{\textbf{\# of}\\\textbf{papers}} & \makecell{\textbf{\# of}\\\textbf{references}} & 
 \makecell{\textbf{\# of}\\\textbf{papers}} & \makecell{\textbf{\# of}\\\textbf{references}} & 
 \makecell{\textbf{\# of shared}\\\textbf{references}} & \makecell{\textbf{\% of shared}\\\textbf{references}} \\
\midrule
14 & 271 & 12,912 & 232,934 & 130 & 2,393 & \textbf{2,206} & \textbf{92.20\%} \\
15 & 271 & 12,912 & 232,934 & 26  & 470   & \textbf{27} & \textbf{5.70\%} \\
\bottomrule
\end{tabular}
\end{table}

\subsubsection{Disambiguation by Label Spreading}\label{subsec5}

Label propagation is a graph-based semi-supervised learning method that has proven effective in domains with scarce labels. It has been widely applied in tasks ranging from natural language processing (e.g. cross-lingual part-of-speech tagging) to computer vision (3D point cloud classification) and network analysis~\cite{sattari2018spreading, Zinilli2025elements}. In particular, label propagation is popular for community detection in large dynamic social networks due to its efficiency~\cite{sattari2018spreading}. In the scientometric domain, label propagation techniques have been employed for author name disambiguation, where a small set of labeled author identities can be propagated through an authorship network. For instance, a recent study constructed a graph of co-authorships and used label propagation to cluster papers by author; this semi-supervised approach achieved high accuracy in the Astrophysics Data System when distinguishing publications of different authors with the same name~\cite{mihaljevic2021}.

In this work, we employ a similar technique to propagate information related to scientific fields. The underlying idea is that, although Scopus authors are not explicitly associated with scientific fields from the above taxonomy (Scientific Areas, Groups, Recruitment Fields or Academic Disciplines), such information can be inferred by propagating the areas of their neighbors in the co-author network, starting from a limited sample of labeled seed nodes. Corresponding entries across sources whose scientific fields also align can then be considered as matching authors.

We consider a graph $G$ where nodes represent Scopus authors, edges indicate co-authorship, and some nodes are labeled with their scientific field. Depending on the desired level of granularity, scientific fields can correspond to Scientific Areas, Groups, Recruitment Fields, or Academic Disciplines. This labeling information is available for a limited number of nodes in CercaUniversit\`{a} whose Scopus AUIDs are known. We propagate labels by employing a Label Spreading approach~\cite{Zhou2004} to infer the label of unlabeled nodes. We then use the inferred label to decide whether the scopus author corresponds to a given CercaUniversit\`{a} author. Next, we discuss the label spreading algorithm.

Let \(W\in\mathbb{R}^{n\times n}\) be a symmetric non‑negative adjacency (similarity) matrix of $G$ with \(n\) vertices and \(C\) classes and $D$ its degree matrix, a diagonal matrix whose diagonal entries are the node degrees. We define the symmetrically normalized similarity
\(S = D^{-1/2} W D^{-1/2}\).
The initial label information is encoded as a one‑hot matrix
\(Y\in\{0,1\}^{n\times C}\) whose \((i,c)\) entry equals 1 if vertex \(i\) is labeled with class \(c\), 0 otherwise.
Label Spreading~\cite{Zhou2004} finds a soft label matrix
\(F\in\mathbb{R}^{n\times C}\) by minimizing

\[
\min_{F}\;
\frac12\sum_{i,j=0}^n S_{ij}\,\lVert F_i-F_j\rVert^2
\;+\;
\lambda \sum_{i=1}^{n}\lVert F_i-Y_i\rVert^2
\qquad
\mbox{with } \lambda = \frac{1-\alpha}{\alpha}
\]

which trades off graph smoothness against fidelity to the known labels, based on a parameter $\alpha$, where $0 < \alpha < 1$.
In practice the solution is obtained through the fixed‑point iteration~\cite{Zhou2004}:

\[
F^{(t+1)} = \alpha\,S\,F^{(t)} + (1-\alpha)\,Y,
\quad
F^{(0)} = Y
\]

The sequence \(\{F^{(t)}\}_{t\ge0}\) converges linearly to the closed‑form limit
\(F^\star = (1-\alpha)(I-\alpha S)^{-1}Y\),
which is the unique minimizer of the objective above.

In our experiments we set $\alpha=0.2$ and stop when
\(\lVert F^{(t+1)}-F^{(t)}\rVert_{\infty}\le0.001\) or after a maximum of 30 iterations.

\subsubsection{LLM disambiguation}\label{subsubsec1}
In this scenario, the author disambiguation task is addressed using the Llama 3.1 70B Instruct model\footnote{\url{https://huggingface.co/meta-llama/Llama-3.1-70B-Instruct}}, employed in zero-shot inference mode.
For each ambiguous case, we assess which types of bibliographic metadata are most informative for the disambiguation task. Specifically, we test three input configurations for each candidate author profile retrieved from Scopus: (i) keywords only, (ii) keywords and titles, and (iii) keywords, titles and abstracts.
Each input configuration provides a different level of semantic richness. Keywords capture the general subject area of the research; titles introduce syntactic and semantic features related to research focus; abstracts provide richer contextual information, although their verbosity can introduce redundancy or ambiguity. This comparison allows us to understand the role of each metadata combination in guiding the disambiguation process.
In all configurations, we also include author-related information such as given name, family name, initials, full name, and affiliations. These are combined with a chronologically ordered list of publications, where each entry provides the publication year and the corresponding metadata according to the tested configuration.
For this preliminary evaluation, we limit the input to a random sample of up to 10 papers per candidate author to manage computational resources given the variability in publication volume.
All the tests are conducted using the same configuration: 
greedy decoding (i.e., \textit{top\_k} = 1), and \textit{max\_length} = 700, while all other parameters are left to their default values. The LLM-based evaluation is performed on a GPU server with three NVIDIA H100 GPUs, each with 80 GB of memory.
Preliminary results showed that the second configuration (keywords and titles) provides the most effective balance between informativeness and efficiency. We therefore selected this setting to perform a more comprehensive evaluation using all available information for each candidate author, rather than a limited sample of 10 publications. This comparison is feasible for this configuration, as keywords and titles are concise enough to allow full inclusion without exceeding computational resource limits.
The full input provided to the model is illustrated in Figures~\ref{fig:prompt} and~\ref{fig:condidate_info}.
\begin{figure}[ht]
\centering
\begin{minipage}{0.95\textwidth}
\begin{tcolorbox}[colback=gray!5, colframe=black!70, title=Disambiguation Task Prompt, label={box:1}]
\textbf{System:} Your task is to evaluate the candidate to determine if they match the specified researcher profile.

\vspace{0.3cm}
\textbf{User:} Is the following candidate a match for the researcher [Name] [Surname], affiliated with [University],
working in the Italian academic field of [SDS Description]?

Here is the candidate: [Candidate's Info]

Based on the provided information, do you believe this candidate is the best match? 

Please respond with ``\texttt{yes}'' or ``\texttt{no}'' and a brief explanation.

Respond only in JSON format as follows:

\begin{verbatim}
{
  "cerca_univ_id": [exact input value],
  "scopus_candidate_id": [exact input value],
  "match": "yes" or "no",
  "explanation": "Your explanation here."
}
\end{verbatim}
\end{tcolorbox}
\end{minipage}
\caption{Prompt used for zero-shot author disambiguation.}
\label{fig:prompt}
\end{figure}

\begin{figure}[ht]
\centering
\begin{minipage}{0.95\textwidth}
\begin{tcolorbox}[title=Candidate Information, colback=gray!10, colframe=gray!80!black, fontupper=\ttfamily]
\texttt{Author ID}: \texttt{<candidate\_id>} \\
\texttt{Name}: \texttt{<full\_name>} \\
\texttt{Surname}: \texttt{<surname>} \\
\texttt{Initials}: \texttt{<initials>} \\
\texttt{Affiliations}: \texttt{<affiliation1>; <affiliation2>} \\[5pt]
\texttt{Publications (chronological order)}:\\
\texttt{- [year A] Title A - Keywords: keyword1, keyword2, ...}\\
\texttt{- [year B] Title B - Keywords: keyword1, keyword2, ...}
\end{tcolorbox}
\end{minipage}
\caption{Example of candidate's information.}
\label{fig:condidate_info}
\end{figure}

\subsubsection{LLM-enriched disambiguation}\label{subsec6}
To integrate semantic and structural dimensions, we design a hybrid pipeline that builds on the LLM-based keywords and titles configuration (cf. Section~\ref{subsubsec1}) and incorporates Label Spreading (cf. Section~\ref{subsec5}) and Bibliographic Coupling (cf. Section~\ref{subsec4}).

The rationale is that, while bibliographic metadata provide semantic features about the research focus of an author, relational information derived from co-authorship and citation networks offers complementary evidence. We therefore test whether embedding such features into the LLM input improves its ability to distinguish between candidate author profiles.

The enrichment strategy combines three complementary modules. The LLM-based component extracts semantic features from scholarly textual metadata, Label Spreading propagates academic information through the co-authorship network, and Bibliographic Coupling exploits shared citations to characterize authors within the citation network. By unifying these sources of evidence, the methodology leverages their respective strengths to achieve a more robust and accurate disambiguation process.

We consider three enrichment variants: (i) Label Spreading only, where the classification derived from the co-authorship network is appended to the textual descriptors; (ii) Bibliographic Coupling only, where the percentage of shared citations is included as an additional feature; and (iii) the combined configuration, where both structural sources are jointly integrated. In all cases, the enriched information is merged with the bibliographic metadata and provided to the LLM in the form of an extended prompt. This allows the model to reason not only on the semantic content of publications but also on the author’s position within the academic network.

The design of this experiment aims to evaluate the complementarity of semantic and structural signals. By systematically testing each enrichment strategy, we can assess whether network-based evidence contributes unique information or instead overlaps with patterns already captured by the LLM. Figure~\ref{fig:prompt_hybrid} illustrates the structure of the enriched input used in this setting.

\subsubsection{LEAD - LLM-enhanced Engine for Author Disambiguation}\label{subsec7}

The LEAD (LLM-enhanced Engine for Author Disambiguation) methodology builds on the insights gained from the enrichment experiments. Unlike the LLM-enriched setting, where the LLM is applied to all candidate cases with prompts augmented by network-based information, LEAD adopts a selective pipeline designed to maximize robustness while controlling computational costs.

The pipeline combines the strengths of all three modules—an LLM-based component, Label Spreading, and Bibliographic Coupling—but organizes them in a two-stage process. First, Bibliographic Coupling is applied to all candidate pairs and serves as the primary disambiguation method, as it consistently provided the best overall performance in our experiments. Second, a reliability threshold is defined on the overlap score: when the value falls below this threshold, the pair is further evaluated by the LLM.
The prompt employed in LEAD corresponds to that of the LLM-enriched configuration (Figure~\ref{fig:prompt_hybrid}), but it is applied only to the hard cases identified in the first stage.
For these selected cases, the LLM is not limited to textual metadata (keywords and titles) but, as in the LLM-enriched setting, also receives structural information derived from the other modules. Specifically, the enriched prompt includes both data on Bibliographic Coupling and the academic classification inferred from the co-authorship network by Label Spreading. This ensures that the model always has access to the same set of semantic and structural signals, but its application is restricted to cases where the baseline method is uncertain.

By concentrating the LLM’s capacity on the most ambiguous instances, LEAD preserves the efficiency of Bibliographic Coupling while mitigating its weaknesses in borderline cases. Compared to a full-scale LLM-enriched approach, the number of model calls is significantly reduced, keeping inference time within practical limits without sacrificing disambiguation accuracy. 

\begin{figure}[ht]
\centering
\begin{minipage}{0.95\textwidth}
\begin{tcolorbox}[colback=gray!5, colframe=black!70, title=Hybrid Disambiguation Prompt, label={box:hybrid}]
\textbf{System:} Your task is to evaluate the candidate to determine if they match the specified researcher profile.\\

\vspace{0.3cm}
\textbf{User:} Is the following candidate a match for the researcher [Name] [Surname], affiliated with [University],
working in the Italian academic field of [AD Description]?

You also have additional information obtained through automated methods, which may contain inaccuracies.
Please use this information critically to support your evaluation:

\begin{enumerate}
    \item An analysis of citations through a citation network has shown that [overlap\%] of the citations in the candidate's [n\_papers] papers (which cite a total of [n\_cited] papers) align with those of the academic community, with [n\_matches] relevant citations found in the network.
    \item A method based on a co-authorship network has predicted the following academic classification:
    \vspace{-0.1cm}
    \begin{itemize}
        \item Recruitment field: [AD] - [AD\_label]
        \item Macro recruitment field: [Group] - [Group\_label]
        \item Area: [area] - [Scientific\_Area\_label]
    \end{itemize}
\end{enumerate}

Here is the candidate: [Candidate's Info]

Based on the provided information, do you believe this candidate is the best match? 

Please respond with ``\texttt{yes}'' or ``\texttt{no}'' and a brief explanation.

Respond only in JSON format as follows:

\begin{verbatim}
{
  "cerca_univ_id": [exact input value],
  "scopus_candidate_id": [exact input value],
  "match": "yes" or "no",
  "explanation": "Your explanation here."
}
\end{verbatim}
\end{tcolorbox}
\end{minipage}
\caption{Prompt used in the hybrid approach: the LLM receives both the candidate's textual metadata and additional evidence from Bibliographic Coupling and Label Spreading.}
\label{fig:prompt_hybrid}
\end{figure}


\section{Results}\label{sec4}
This section reports the experimental results of the proposed disambiguation pipeline. The evaluation relies on binary classification metrics - Precision, Recall, F1, and Accuracy - chosen to reflect the formulation of our task. Specifically, the evaluation focuses on hard cases, i.e., ambiguous candidates that are particularly challenging to resolve. Rather than producing ranked lists of alternatives, the system must make an individual binary choice for each candidate, deciding whether it matches the target researcher. Within this framework, Precision expresses the proportion of predicted matches that are correct, Recall measures the proportion of true matches successfully identified, the F1 balances these two aspects, and Accuracy captures the overall proportion of correctly classified candidates.

\subsection{Overall performance}\label{overall-results}

\begin{table}[h]
\caption{Overall performance comparison of disambiguation methods based on binary classification metrics. Precision, Recall, F1, and Accuracy are reported as percentages. The best result for each metric is highlighted in bold, while the second-best is underlined.}\label{overall_performance}
\begin{tabular*}{\textwidth}{@{\extracolsep\fill}lccccr}
\toprule
\textbf{Method} & \textbf{Precision} & \textbf{Recall} & \textbf{F1} & \textbf{Accuracy} & \textbf{Time (s)}\\
\midrule
Label Spreading & 85.0 & 87.8 & 86.3 & 81.8 & 2.432\\
Bibliographic Coupling & 96.6 & \underline{92.9} & \underline{94.7} & \underline{93.2} & 1.883 \\
LLM  & \textbf{97.7} & 89.8 & 93.6 & 92.1 & 5381.0 \\ 
LLM-enriched & \underline{97.0} & 92.4 & \underline{94.7} & \underline{93.2} & 4602.0 \\
LEAD & 96.2 & \textbf{97.2} &   \textbf{96.7} & \textbf{95.7} & 1842.0 \\
\botrule
\end{tabular*}
\end{table}

Table~\ref{overall_performance} provides a comparative overview of the disambiguation methods presented in Section~\ref{subsec3}, each evaluated under the conditions that maximized its performance. Specifically, results correspond to: (i) Label Spreading at the Scientific Areas level; (ii) Bibliographic Coupling on the complete 8-year citation network, where matches are accepted only when their overlap score exceeds 0.15; (iii) the LLM-based approach with the keywords and titles setting applied to all available papers per author; (iv) the LLM-enriched configuration adding only Bibliographic Coupling information; and (v) LEAD applied only to hard cases identified by Bibliographic Coupling with an overlap score below 0.15 on the complete network.
These results provide a comprehensive comparison of the five disambiguation methods, highlighting the trade-offs between performance and efficiency.
Label Spreading and Bibliographic Coupling are the most lightweight approaches, requiring only 2.432 and 1.833 seconds, respectively, to complete the disambiguation task. 

The Label Spreading results indicate a balanced outcome, with precision of 85.0\% and recall of 87.8\%, yielding an F1 of 86.3\% and an accuracy of 81.8\%. This performance highlights its role as an efficient baseline, capable of providing rapid results, though less effective than the other approaches.


Bibliographic Coupling demonstrates strong performance, combining very high precision of 96.6\% with the second-best recall of 92.9\%. This combination results in an F1 of 94.7\% and an accuracy of 93.2\%, making it the strongest single-source method in terms of both precision and recall. Its effectiveness comes from leveraging structural evidence in the citation network, making it possible to capture true matches extensively while retaining high precision.

The LLM-based approach achieves the highest precision of 97.7\%. However, its recall reaches 89.8\%, below the 92.9\% of Bibliographic Coupling, resulting in an F1 of 93.6\% and an accuracy of 92.1\%. The gain in precision comes at a considerable computational cost, with an inference time of 5381 seconds. These results indicate that the LLM is particularly effective at identifying highly reliable matches, but its selectivity leads to a loss of valid correspondences.

The LLM-enriched configuration with Bibliographic Coupling achieves the same F1 of 94.7\% and accuracy of 93.2\% as Bibliographic Coupling. Precision increases to 97.0\%, while recall slightly decreases to 92.4\%. This demonstrates that the structural signal from the citation network already captures most of the discriminative information, making the semantic addition computationally costly but ultimately redundant.

The proposed hybrid method, LEAD, emerges as the most effective approach. By selectively applying the LLM only to ambiguous cases identified through Bibliographic Coupling, it achieves the highest recall at 97.2\% while maintaining strong precision of 96.2\%, resulting in the best overall performance with an F1 of 96.7\% and an accuracy of 95.7\%. Its inference time of 1842 seconds is significantly lower than both the full LLM approaches and the LLM-enriched variants. These results demonstrate the strength of a selective hybrid strategy, which maximizes effectiveness and surpasses both single-source and enriched methods while keeping the computational cost at a level that is reasonable when top disambiguation performance is the priority.

\begin{table}[t]
\centering
\caption{An example from the gold standard (Record~ID~49). Bibliographic Coupling (BC) fails to disambiguate the two Scopus profiles, assigning scores below the acceptance threshold and predicting \texttt{No} for both. LEAD, instead, predicts \texttt{Yes} for both candidates, correctly identifying them as the same author.}
\label{example_table}
\small
\setlength{\tabcolsep}{3pt} 
\begin{tabular}{lccccccc}
\toprule
\textbf{Record ID} & \textbf{AUID} & \textbf{Papers} & \textbf{References} & \textbf{Shared refs} & \textbf{Shared (\%)} & \textbf{BC} & \textbf{LEAD} \\
\midrule
49 & 6603258864  & 14 & 597 & 23 & 3.9 & No  & Yes \\
49 & 57208832161 &  1 &  60 &  1  & 1.7 & No  & Yes \\
\bottomrule
\end{tabular}
\end{table}

To further illustrate these results, Table~\ref{example_table} presents a representative case from the gold standard. Record~ID~49 involves a CercaUniversità academic that is associated with two Scopus profiles. One profile contains a single paper with 60 references, whereas the other includes 14 papers with 597 references. Using the optimal threshold of 0.15, Bibliographic Coupling fails to match them, yielding overlap scores of only 3.9\% and 1.7\%. In such sparse conditions, structural evidence alone is not sufficient for reliable disambiguation. Within LEAD, the case is treated as hard and processed by the LLM component. The LLM reasoning over textual metadata, combined with Label Spreading and Bibliographic Coupling results, correctly matches both profiles independently.
This example highlights LEAD’s ability to resolve cases where structural evidence is insufficient by strategically leveraging semantic information. Such an integration ensures disambiguation performance that consistently exceeds the capabilities of individual methods.

\subsection{Performances of LLM-based approaches}\label{llm-results}
We evaluated the LLM-based approach introduced in Section~\ref{subsubsec1} on the dataset described in Section~\ref{subsec2}, considering three metadata configurations: only keywords; keywords and titles; and keywords, titles, and abstracts. As reported in Table~\ref{llm-comparison}, relying solely on keywords already provides a strong baseline, with an F1 of 90.0\% and an accuracy of 87.4\%. Adding titles improves all metrics, raising F1 to 91.6\%, accuracy to 89.7\%, and recall to 86.3\%, which is the highest among the three settings. Including abstracts further increases precision to 98.3\% and yields the best overall F1 of 91.7\% and accuracy of 89.9\%, although recall decreases slightly compared to the {\em keywords and titles} configuration. These results suggest that while abstracts add information, they may also introduce redundancy or noise, limiting their discriminative value. Overall, the {\em keywords and titles} configuration offers the most favorable compromise between precision and recall, with abstracts providing only marginal additional gains.

\begin{table}[h]
\caption{Performance of LLM-based disambiguation approaches. Results are reported for the three metadata configurations (keywords only; keywords and titles; keywords, titles and abstracts), evaluated both with a limit of 10 papers per author and with all available publications. The lower part of the table shows the results of the LLM-enriched variants. Precision, Recall, F1, and Accuracy are reported in percentage (\%). The best-performing configurations are highlighted in gray.}
\label{llm-comparison}
\small
\setlength{\tabcolsep}{1pt} 
\centering
\begin{tabular}{lccccc}
\toprule
\textbf{Dataset} & \textbf{Precision} & \textbf{Recall} & \textbf{F1} & \textbf{Accuracy} & \textbf{Time (s)} \\
\midrule
Only keywords (10) & 96.2 & 83.9 & 90.0 & 87.4 & 4720.0 \\
Keywords and titles (10) & 97.7 & 86.3 & 91.6 & 89.7 & 4807.0 \\
Keywords, titles and abstracts (10) & \underline{98.3} & 86.0 & 91.7 & 89.9 & 5051.0 \\
\rowcolor{gray!15}
Keywords and titles  & 97.7 & \underline{89.8} & \underline{93.6} & \underline{92.1} & 5381.0 \\
\midrule
\multicolumn{6}{l}{\textbf{LLM-enriched (Keywords and titles)}} \\ 
\quad + Label Spreading & \textbf{98.8} & 85.0 & 91.4 & 89.6 & 5097.0 \\
\rowcolor{gray!15}
\quad + Bibliographic Coupling & 97.0 & \textbf{92.4} & \textbf{94.7} & \textbf{93.2} & 4602.0 \\
\quad + Label Spreading + Bibliographic Coupling & \textbf{98.8} & 84.2 & 90.9 & 89.1 & 5108.0 \\
\bottomrule
\end{tabular} 
\end{table}

To further evaluate the robustness of the LLM-based approach, we focus on the {\em keywords and titles} metadata configuration, comparing its performance when restricted to a subset of 10 papers per author (Keywords and titles (10)) and when applied to the complete dataset (Keywords and titles). Leveraging all papers rather than the 10-paper subset increases recall from 86.0\% to 89.8\%, F1 from 91.6\% to 93.6\%, and accuracy from 89.7\% to 92.1\%, while keeping unchanged the value of precision at 97.7\%.  
These performance gains come at the cost of longer inference time (5381 s vs. 4807 s), representing a moderate increase in computational demand. The comparison indicates that access to a more complete dataset significantly enhances the model’s ability to generalize, with only a reasonable trade-off in processing time.

Finally, we explored LLM-enriched variants of the \textit{keywords and titles} configuration, reported in the lower part of Table~\ref{llm-comparison}. When combined with Label Spreading, the approach achieved the highest precision of 98.8\%, but recall dropped to 85.0\%, resulting in an F1 of 91.4\%. In contrast, Bibliographic Coupling offered the most favorable balance, raising recall to 92.4\% and yielding the best overall F1 of 94.7\% and accuracy of 93.2\%, while also reducing inference time to 4602 seconds. This makes it the most cost-effective enrichment strategy. The combined variant with both Label Spreading and Bibliographic Coupling did not improve over Bibliographic Coupling alone, as recall decreased to 84.2\% and F1 to 90.9\%.

\subsection{Performances for Label Spreading}
Table~\ref{labelspreading-comparison} reports the performance of the Label Spreading disambiguation method across different levels of taxonomic granularity. At the most detailed level of Recruitment Fields, the method reaches the highest precision of 89.4\%, confirming that positive matches are highly reliable. Recall, however, remains lower at 74.9\%, which limits the ability to recover all true correspondences and results in an F1 of 81.5\% and an accuracy of 77.9\%. This outcome reflects the limitations of fine-grained categories, since many small and sparsely populated fields constrain the spread of labels through the co-authorship network and reduce the effectiveness of propagation. The resulting assignments are therefore precise but only partially comprehensive.

Scientific Areas, by contrast, achieve the highest recall of 87.6\% together with the best F1 of 86.3\% and accuracy of 81.8\%. Precision decreases to 85.0\%, which reflects a modest risk of grouping together authors from related but distinct fields. Nevertheless, the broader categories enable labels to propagate more widely through the network, allowing the method to identify a larger fraction of true correspondences. In this setting, the gain in recall outweighs the small reduction in precision, producing the best overall performance.

Recruitment Field Groups fall in between, with precision of 89.1\%, recall of 76.4\%, F1 of 82.2\%, and accuracy of 78.5\%. This intermediate level provides a compromise between the narrow but highly precise Recruitment Fields and the broader but less specific Scientific Areas.

These results show that the effectiveness of Label Spreading depends strongly on the chosen level of granularity. Fine-grained taxonomies provide highly reliable but less comprehensive matches, broad categories maximize the identification of correspondences at the cost of reduced specificity, and intermediate levels provide the most balanced trade-off between the two.

\begin{table}[h]
\caption{Performance of Label Spreading disambiguation across Recruitment Fields (RF), Recruitment Fields Groups, and Scientific Areas (SA). Precision, Recall, F1, and Accuracy are reported as percentage (\%). The best-performing configurations are highlighted in gray.}\label{labelspreading-comparison}
\begin{tabularx}{\textwidth}{l *{4}{>{\centering\arraybackslash}X}}
\toprule
\textbf{Classification Level} & \textbf{Precision} & \textbf{Recall} & \textbf{F1} & \textbf{Accuracy} \\
\midrule
Recruitment Fields & \textbf{89.4} & 74.9 & 81.5 & 77.9 \\
Recruitment Fields Groups & 89.1 & 76.4 & 82.2 & 78.5 \\
\rowcolor{gray!15}
Scientific Areas & 85.0 & \textbf{87.6} & \textbf{86.3} & \textbf{81.8} \\
\botrule
\end{tabularx}
\end{table}



\subsection{Performances for Bibliographic Coupling}

\begin{table}[ht]
\centering
\caption{Performance of Bibliographic Coupling by threshold on two networks. Precision, Recall, F1, and Accuracy are
reported as percentage (\%). The best-performing configurations are highlighted in gray.}
\label{tab:cit_overlap_pct}
\small
\setlength{\tabcolsep}{3.5pt} 
\begin{tabular}{lcccccccc}
\toprule
 & \multicolumn{4}{c}{\textbf{4-Years Network (2020--2023)}} & \multicolumn{4}{c}{\textbf{8-Years Network (2016--2023)}} \\
\cmidrule(lr){2-5}\cmidrule(lr){6-9}
\textbf{Threshold} & \textbf{Precision} & \textbf{Recall} & \textbf{F1} & \textbf{Accuracy}
                   & \textbf{Precision} & \textbf{Recall} & \textbf{F1} & \textbf{Accuracy} \\
\midrule
0.25 & 98.8 & 82.7 & 90.1 & 88.1 & \textbf{99.1} & 86.0 & 92.1 & 90.5 \\
0.20 & 97.7 & 85.8 & 91.4 & 89.4 & 97.8 & 89.1 & 93.0 & 91.6 \\
0.15 & 97.3 & 89.8 & 93.4 & 91.7 
     & \cellcolor{gray!15}96.6 & \cellcolor{gray!15}92.9 & \cellcolor{gray!15}\textbf{94.7} & \cellcolor{gray!15}\textbf{93.2} \\
0.10 & 95.0 & 91.9 & 93.4 & 91.6 & 94.7 & \textbf{94.4} & 94.5 & 92.9 \\
\bottomrule
\end{tabular}
\end{table}


For the Bibliographic Coupling disambiguation, we evaluate two citation networks with different temporal spans: a four-year network (2020--2023) and an extended eight-year network (2016--2023). Performance is assessed at four threshold levels (0.25, 0.20, 0.15, and 0.10) to capture precision–recall trade-offs and identify optimal configurations (Table~\ref{tab:cit_overlap_pct}).

Across all thresholds, the extended network consistently outperforms the four-year network. The gains are most evident when comparing the two networks at the same threshold, with recall increasing by up to 3.3 percentage points and F1 improving by as much as 2.2 points. These improvements confirm that broader temporal coverage increases the density of shared references, reduces network sparsity, and captures more stable scholarly connections, thereby strengthening the structural evidence available for validation.  

Precision remains extremely high at the highest threshold, reaching 98.8\% for the four-year network and 99.1\% for the extended network at 0.25. As thresholds become more permissive, recall rises substantially while precision stays consistently above 94\%, showing that the method maintains robustness under different operational settings.

The most favorable configuration is obtained at threshold 0.15 on the extended network, where precision of 96.6\% and recall of 92.9\% yield the best overall performance, with an F1 of 94.7\% and accuracy of 93.2\%. This setting illustrates how temporal extension and calibrated thresholding together provide the strongest results, balancing conservativeness with inclusiveness and avoiding invalid matches.
\subsection{Parameter Sensitivity Analysis}

\begin{table}[ht]
\centering
\caption{Parameter sensitivity analysis of LEAD across different thresholds (0.25, 0.20, 0.15, 0.10), with and without Label Spreading (LS) and Bibliographic Coupling (BC). Results are reported in terms of Precision, Recall, F1, Accuracy, and computational time. The best-performing configurations are highlighted in gray.}
\label{tab:ablation}
\begin{tabular}{lllcccccc}
\toprule
\textbf{Method} & \textbf{LS} & \textbf{BC} & \textbf{Threshold} & \textbf{Precision} & \textbf{Recall} & \textbf{F1} & \textbf{Accuracy} & \textbf{Time (s)} \\
\midrule
\multirow{13}{*}{LEAD} 
  & No  & Yes & \multirow{3}{*}{0.25} & \cellcolor{gray!15}94.4 & \cellcolor{gray!15}98.4 & \cellcolor{gray!15}96.4 & \cellcolor{gray!15}95.2 & \cellcolor{gray!15}1895 \\
  & Yes & No  &                       &   97.9  &   93.1  &   95.4  &   94.2  &   2116  \\
  & Yes & Yes &                       & \textbf{98.7} & 93.1 & 95.8 & 94.7 & 2170 \\
\cmidrule(l){2-9}
  & No  & Yes & \multirow{3}{*}{0.20} & 93.3 & 98.5 & 95.8 & 94.4 & 1766 \\
  & Yes & No  &                       &   96.6  &   94.4  &   95.5  &   94.2  &   1944  \\
  & Yes & Yes &                       & \cellcolor{gray!15}97.4 & \cellcolor{gray!15}94.7 & \cellcolor{gray!15}96.0 & \cellcolor{gray!15}94.9 & \cellcolor{gray!15}2007 \\
\cmidrule(l){2-9}
  & No  & Yes & \multirow{3}{*}{0.15} & 92.4 & 98.7 & 95.5 & 93.9 & 1628 \\
  & Yes & No  &                       &   95.7  &   96.7  &   96.2  &   95.0  &   1778  \\
  & Yes & Yes &                       & \cellcolor{gray!15}96.2 & \cellcolor{gray!15}97.2 & \cellcolor{gray!15}\textbf{96.7} & \cellcolor{gray!15}\textbf{95.7} & \cellcolor{gray!15}1842 \\
\cmidrule(l){2-9}
  & No  & Yes & \multirow{3}{*}{0.10} & 90.9 & \textbf{98.7} & 94.7 & 92.7 & 1528 \\
  & Yes & No  &                       &   93.9  &   97.5  &   95.6  &   94.2  &   1675  \\
  & Yes & Yes &                       & \cellcolor{gray!15}94.3 & \cellcolor{gray!15}97.7 & \cellcolor{gray!15}96.0 & \cellcolor{gray!15}94.2 & \cellcolor{gray!15}1721 \\
\bottomrule
\end{tabular}
\end{table}

The parameter sensitivity analysis presented in Table~\ref{tab:ablation} provides insights into the performance of LEAD under varying threshold configurations and component combinations. The results reveal consistent trade-offs between precision and recall that depend on threshold selection.  

At the highest threshold of 0.25, the method operates conservatively, prioritizing precision over coverage. When both Label Spreading (LS) and Bibliographic Coupling (BC) are incorporated, LEAD achieves its maximum precision of 98.7\%, effectively minimizing false positives. This comes at the expense of recall, which falls to 93.1\%.  

At the lowest threshold of 0.10, with only BC the method achieves maximum recall of 98.7\% but precision drops to 90.9\%. With only LS, precision improves to 93.9\% while recall remains high at 97.5\%. The combined configuration achieves the most balanced outcome, with 94.3\% precision and 97.7\% recall, showing that LS mitigates false positives without substantially reducing coverage.  

Intermediate thresholds produce more balanced outcomes. At 0.20, the BC-only configuration exhibits precision of 93.3\% and recall of 98.5\%, while LS alone achieves 96.6\% precision and 94.4\% recall. Their integration delivers precision of 97.4\% and recall of 94.7\%, corresponding to an F1-score of 96.0\% and accuracy of 94.9\%.  

The most favorable balance is observed at 0.15, where the joint use of LS and BC provides the highest F1-score of 96.7\% and peak accuracy of 95.7\%. At this threshold, precision (96.2\%) and recall (97.2\%) are closely aligned, reflecting effective control of both false positives and false negatives. Considered individually, LS consistently favors precision at the cost of recall, whereas BC favors recall but admits more false positives. Their combination demonstrates that complementary structural information strengthens the robustness of LEAD by compensating for the limitations of each component in isolation.

The analysis also highlights computational trade-offs across different threshold and component configurations. Processing times vary both with threshold selection and component combinations, ranging from 1528 seconds for the BC-only configuration at threshold 0.10 to 2170 seconds when both LS and BC are incorporated at threshold 0.25. The optimal configuration at threshold 0.15 requires 1842 seconds, representing a reasonable computational investment given its superior F1-score and accuracy. Across all thresholds, LS consistently adds computational overhead compared to BC-only configurations, with the combined approach requiring the most processing time. This additional cost is justified by the performance improvements achieved, as the integration of both components reduces false positives while preserving high recall, thereby ensuring more reliable and stable disambiguation outcomes.  


\section{Conclusions and future work}\label{sec6}
This study addressed the challenging problem of cross-source author name disambiguation, specifically linking records from CercaUniversit\`{a} (i.e. the Italian academic database) with Scopus author profiles. We introduced LEAD, a hybrid two-stage approach that strategically combines Bibliographic Coupling, Label Spreading, and Large Language Models. Evaluated on a manually curated dataset of highly ambiguous cases, LEAD consistently outperformed individual methods while requiring substantially less computational time than full LLM-based approaches.
This study highlights that integrating different sources of evidence provides more reliable results than single-method solutions. By applying LLMs only to the most ambiguous cases, the approach achieves both robustness and efficiency. Such selective integration offers an effective strategy for designing disambiguation systems that balance accuracy with efficiency.
Nonetheless, some challenging cases remain unresolved. These could be addressed through limited human-in-the-loop post-processing or through agent-based AI approaches capable of autonomously querying external sources such as departmental websites, institutional profiles, or academic social networks. Such extensions would enable richer contextual reasoning and provide evidence beyond traditional bibliographic metadata.
While these approaches address immediate disambiguation challenges, broader systemic issues emerge when considering large-scale deployment. Future research should focus on the temporal robustness of disambiguation systems. Although LEAD performs strongly on static datasets, bibliographic databases such as Scopus are dynamic, with author profiles continuously updated, merged, or deleted. Developing adaptive mechanisms will be essential to preserve accuracy and ensure the reproducibility of bibliometric analyses over time. Extending LEAD to other bibliographic platforms would further broaden its applicability and reinforce its role within integrated research information systems.
In conclusion, author disambiguation remains a complex and evolving challenge, but LEAD provides a solid foundation for future systems designed to support the long-term needs of research information management and bibliometric analysis.

\section*{Declarations}

\subsection*{Funding}

This work was supported by FOSSR (Fostering Open Science in Social Science Research), funded by the European Union - NextGenerationEU under NRRP Grant agreement n. MUR IR0000008. The content of this article reflects only the author's view. The European Commission and MUR are not responsible for any use that may be made of the information it contains.

\section*{Ethics declarations}

\subsection*{Conflict of interest}

The authors have no Conflict of interest to declare that are relevant to the content of this article.

\bibliography{sn-bibliography}

\end{document}